\documentclass[aip,preprint]{revtex4-1}
\usepackage{graphicx}
\draft 

\begin{document}


\title{High-pressure neutron study of the morphotropic PZT: phase transitions in a two-phase system} 



\author{J. Frantti}
\email[]{johannes.frantti@aalto.fi}
\author{Y. Fujioka}
\affiliation{Aalto University School of Science, Department of Applied Physics, FI-00076 Aalto, Finland}
\author{J. Zhang}
\author{S. Wang}
\author{S. C. Vogel}
\affiliation{Los Alamos Neutron Science Center, Los Alamos National Laboratory, Los Alamos, New Mexico 87545}
\author{R. M. Nieminen}
\affiliation{Aalto University School of Science, Department of Applied Physics, FI-00076 Aalto, Finland}
\author{A. M. Asiri}
\affiliation{Chemistry Department, Faculty of Science, King Abdulaziz University, P.O. Box 80203, Jeddah 21589, Saudi Arabia and 
Center of Excellence for Advanced Materials Research , King Abdulaziz University, P.O. Box 80203, Jeddah 21589, Saudi Arabia}
\author{Y. Zhao}
\affiliation{Los Alamos Neutron Science Center, Los Alamos National Laboratory, Los Alamos, New Mexico 87545}
\author{A. Y. Obaid}
\affiliation{Chemistry Department, Faculty of Science, King Abdulaziz University, P.O. Box 80203, Jeddah 21589, Saudi Arabia and 
Center of Excellence for Advanced Materials Research , King Abdulaziz University, P.O. Box 80203, Jeddah 21589, Saudi Arabia}
\date{\today}

\begin{abstract}
In piezoelectric ceramics the changes in the phase stabilities versus stress and temperature in the vicinity of the phase boundary play a central role. The present study was dedicated to the classical piezoelectric, lead-zirconate-titanate (PZT) ceramic with composition Pb(Zr$_{0.54}$Ti$_{0.46}$)O$_3$ at the Zr-rich side of the morphotropic phase boundary at which both intrinsic and extrinsic contributions to piezoelectricity are significant. The pressure-induced changes in this two-phase (rhombohedral $R3c$+monoclinic $Cm$ at room temperature and $R3c$+$P4mm$ above 1 GPa pressures) system were studied by high-pressure neutron powder diffraction technique. The experiments show that applying pressure favors the $R3c$ phase, whereas the $Cm$ phase transforms continuously to the $P4mm$, which is favored at elevated temperatures due to the competing entropy term. The $Cm\rightarrow R3c$ phase transformation is discontinuous. The transformation contributes to the extrinsic piezoelectricity. An important contribution to the intrinsic piezoelectricity was revealed: a large displacement of the $B$ cations (Zr and Ti) with respect to the oxygen anions is induced by pressure. Above 600 K a phase transition to a cubic phase took place. Balance between the competing terms dictates the curvature of the phase boundary. After high-pressure experiments the amount of rhombohedral phase was larger than initially, suggesting that on the Zr-rich side of the phase boundary the monoclinic phase is metastable.

\end{abstract}


\maketitle 

\section{Introduction}
Piezoelectric lead-zirconate-titanate [Pb(Zr$_x$Ti$_{1-x}$)O$_3$, PZT] solid solution system was developed over 40 years ago yet attempts to understand its properties continue to trigger new studies. A long-lasting view is that when $x$ is approximately 0.52, a first-order phase transition occurs between tetragonal and rhombohedral phases, resulting in two-phase co-existence. The electromechanical properties peak slightly on the rhombohedral side of the phase boundary. In the composition-temperature plane the boundary (commonly called as the morphotropic phase boundary, MPB) is nearly independent of temperature, thus making PZT very practical material for applications \cite{Jaffe}. The commonly offered reasoning for the exceptionally good electromechanical coupling is based on the idea that there are eight (rhombohedral phase) and six (tetragonal phase) spontaneous polarization directions available in the two-phase system so that the system can readily respond to external electric field or stress. 

The space group symmetries given for a disordered solid-solution should be taken as average symmetries from which short-range order deviates. For instance, it has been known for long that Raman scattering data cannot be explained by the average symmetries. The high-temperature cubic phase has no first-order Raman modes yet experiments revealed that spectra collected on PZT above the Curie temperature have rather strong features at energies close to the low-temperature first-order phonon energies. In the case of so-called relaxor ferroelectrics this type of behavior is normal and the frequently offered explanation is that symmetry-lowering defects generate polar nanoregions (see, e.g., refs. \onlinecite{Uwe,Buixaderas,Frantti_PRB_1996}). Also the low-temperature Raman spectra of Ti-rich PZT have many features which are not consistent with the tetragonal symmetry: the twofold degenerate $E$-symmetry modes of the tetragonal PZT were split, indicating that the symmetry is lower than $P4mm$ \cite{Frantti_PRB_1999}. Raman experiments showed that anharmonicity plays a significant role in lead titanate, the anharmonic contribution being increased with increasing temperature \cite{Foster}. The traditional view was modified once high-resolution x-ray synchrotron studies revealed that the phase believed to be tetragonal possesses monoclinic distortion \cite{Noheda} in the vicinity of the MPB. Neutron powder diffraction experiments, able to resolve the monoclinic split \cite{Frantti_JJAP_2000}, ruled out octahedral tilts, and verified the $Cm$ symmetry \cite{Frantti_JJAP_2000,Frantti_PRB_2002}. 

Accurate modeling of the system requires not only the consideration of the unit cell but also crystallographic twins (or ferroelectric domains) and grain boundaries must be taken into account. In piezoelectric ceramics the response to external stress or electric field can be divided into intrinsic and extrinsic contributions \cite{Newnham}. The former is essentially a single crystal response (i.e. is formed by the ion displacements within a primitive cell of the crystal), whereas the latter covers the contribution due to  grain boundaries, preferred orientation or texture of the grains, i.e. ferroelectric domains within the grains, and changes in crystal phase fractions. Since the full model considering contributions from atomic scale up to the macroscopic grain size scale is very complex, experimental studies have commonly been applied to gain deeper insight. 

Non-180$^{\circ}$ domain switching (i.e., contributing to the extrinsic contribution) gives rise to approximately 34\% of the measured $d_{33}$ coefficient of PZT \cite{Jones}. The extrinsic contribution can be larger or smaller if the domain wall motion is respectively made easier or more difficult by doping \cite{Pramanick_JACS_2009_A,Pramanick_JACS_2009_B}. A study of the domain switching showed that the 90$^{\circ}$ domains in single phase tetragonal phase (titanium rich PZT) hardly switch, whereas the domains in the two-phase region switch \cite{Li}. Texture and strain analysis of the ferroelastic behavior of Pb(Zr$_{0.49}$Ti$_{0.51}$)O$_3$ by in-situ neutron diffraction technique showed that the rhombohedral phase plays a significant role in the macroscopic electromechanical behavior of this material \cite{Rogan}. The domain nucleation and domain wall propagation are central factors limiting the speed of ferroelectric polarization switching \cite{Grigoriev_PRL_2008,Grigoriev_PRB_2009}.

An important intrinsic contribution to the piezoelectricity is due to the increase of certain piezoelectric constants once the phase transition is approached. This increase was predicted to be significant in the vicinity of the pressure-induced phase transition in lead titanate \cite{Frantti_JPCB_2007}. The computations carried out for lead titanate further show that it is the competition between two factors which determines the morphotropic phase boundary \cite{Frantti_JPCB_2009}. The first is the oxygen octahedral tilting, favoring the rhombohedral $R3c$ phase, and the second is the entropy, which in the vicinity of the morphotropic phase boundary favors the tetragonal phase above 130 K. If the two factors are in balance over a large temperature range, a steep phase boundary results in the pressure-temperature plane which is desirable for applications. The advantageous feature of the $R3c$ phase is its ability to be compressed efficiently by tilting the oxygen octahedra, in contrast to symmetries prohibiting oxygen octahedral tilting \cite{Frantti_JPCM_2008}.

\begin{figure}[htb!]
\begin{center}
\includegraphics[angle=0,width=13cm]{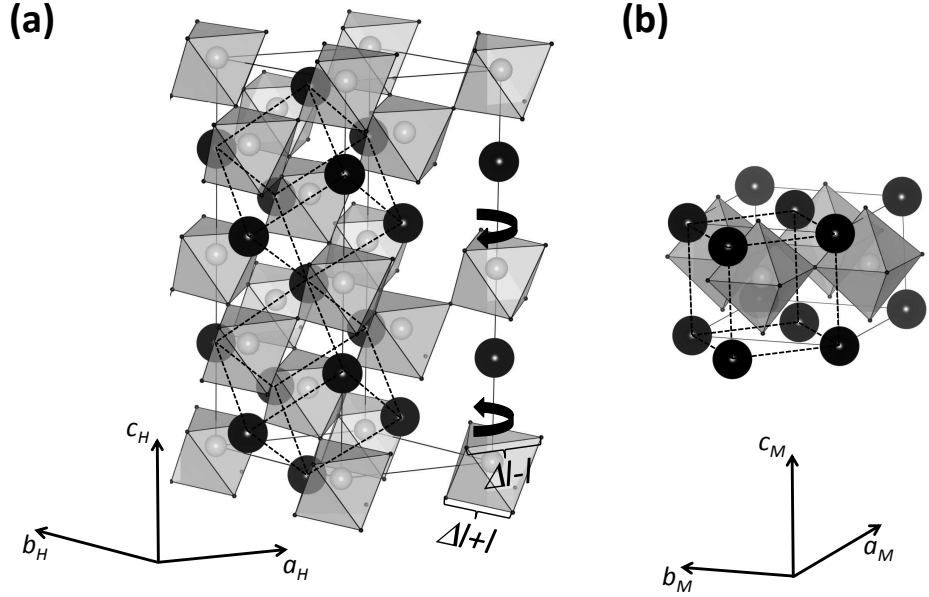}
\caption{\label{Figure1}The $R3c$ phase, whose hexagonal unit cell is shown in panel (a), and the $Cm$ phase, panel (b), behave very differently under applied pressure. The $V_A/V_B$ ratio between the oxygen octahedral and cuboctahedra-volumes of the $R3c$ phase decreases with increasing pressure: the crystal is contracting and thus the $B$ cations (which fit oxygen octahedra tightly) have to take larger relative volume from the total volume (from the cuboctahedra, which has excess of space for Pb) by tilting oxygen octahedra. The symmetry prohibits this mechanism in the $P4mm$ and $Cm$ phases. Density-functional theory computations predict that $P4mm$ has an entropy term benefit at elevated temperatures. Two rhombohedral (corresponding to the $R3m$ phase) pseudocubic cells are shown by dotted lines in panel (a). Due to the octahedral tilting, indicated by arrows, the two cells are not equivalent: the tilting corresponds to the $R3m \rightarrow R3c$ symmetry lowering. The primitive cell of the $Cm$ phase is shown by dotted lines in panel (b). Structure figure was prepared by the VESTA software \cite{VESTA}.}
\end{center}
\end{figure}

We briefly summarize the relationship between the structural parameters and polyhedral tilts and volumes, given in ref. \cite{Thomas}. We follow ref. \onlinecite{Megaw} and parametrize the asymmetric unit of the $R3c$ phase as given in Table \ref{Asym}.
\begin{table}[htb!]
\begin{center}
\caption{\label{Asym}The asymmetric unit of the $R3c$ phase as defined in ref. \onlinecite{Megaw}.}
\begin{tabular}{l l@{   }l@{   }l}
\hline
	& $x$                                      & $y$                              & $z$                      \\
Pb	& $0$                                      & $0$                              & $s+\frac{1}{4}$ \\
Ti/Zr	& $0$                                      & $0$                              & $t$                       \\
O	& $\frac{1}{6}-2e-2d$           & $\frac{1}{3}-4d$        & $\frac{1}{12}$    \\
\hline
\end{tabular}
\end{center}
\end{table}

There is one short and one long O-O octahedral edge length parallel to the hexagonal $ab$-plane, labeled as $l-\Delta l$ and $l+\Delta l$, respectively (see also Fig. \ref{Figure1}). Now, the octahedral tilt angle is given by 
$\tan\omega = 3^{1/2} 4 e$ and the polyhedral volume ratio $V_A/V_B$ is equal to $6K^2\cos^2\omega-1$, where $K$ is given by equation $a = 2 K l \cos\omega$.
The present study focuses on the two-phase, $Cm$ and $R3c$, PZT ceramic material, Pb(Zr$_{0.54}$Ti$_{0.46}$)O$_3$, which has a composition slightly on the Zr-rich side of the morphotropic phase boundary. The main goal was to determine the phase fractions and structural parameters as a function of applied pressure and temperature. Also the question concerning the reversibility of the structural properties of PZT is addressed.

\section{Experimental}
To address the possible homogeneity differences due to the variation in solid-state reaction based sample preparation method lead zirconate-titanate powders were prepared using different starting oxides and sintering conditions. In the first route the PbO, ZrO$_2$ and TiO$_2$ oxides were mechanically mixed in desired proportions, whereas in the second method PbTiO$_3$ and PbZrO$_3$ powders were used as starting chemicals. The phase purity and crystal structure were checked by X-ray powder diffraction and scanning electron microscopy measurements. No significant differences were observed and thus a sample prepared through the latter method was used for the experiments. Samples were annealed by first forming perovskite structure at  1073 K (30 minutes), then increasing the temperature to 1373 K (60 minutes) to improve the sample homogeneity and then cooling the sample first to a stepwise manner to room temperature. Annealing times were kept rather short in order to limit PbO loss.
High-pressure neutron powder diffraction experiments were carried out at the Los Alamos Neutron Scattering Center using the TAP-98 toroidal anvil press \cite{Zhao_HPR_1999,Zhao_2005} set on the high-pressure-preferred orientation (HIPPO) diffractometer \cite{Wenk_2003,Vogel}. Pressure was generated using the high-pressure anvil cells. Sodium chloride was used as a pressure calibrant material. To minimize deviatoric stress built up during room-temperature compression on the polycrystalline sample, all data in our high P-T neutron-diffraction experiment were collected during the cooling cycle from 800 K at each desired loading pressure. Data were collected between 300 and 800 K as a function of pressure. Rietveld refinements were carried out using the program General Structure Analysis System (GSAS) \cite{GSAS} and EXPGUI \cite{EXPGUI}. The pressure was estimated from the reflection positions of the NaCl phase through the equation of state \cite{Decker}. At higher pressures it was necessary to include the reflections from the diamond anvils in the refinement model. The broad hump seen in the background intensity between 2 and 3 \AA{} is due 
to the diffuse scattering from the amorphous zirconium phophate gasket and was modelled using the diffuse scattering option available in the GSAS software.

\section{Results and discussion}
\begin{figure}[htb!]
\begin{center}
\includegraphics[angle=0,width=13cm]{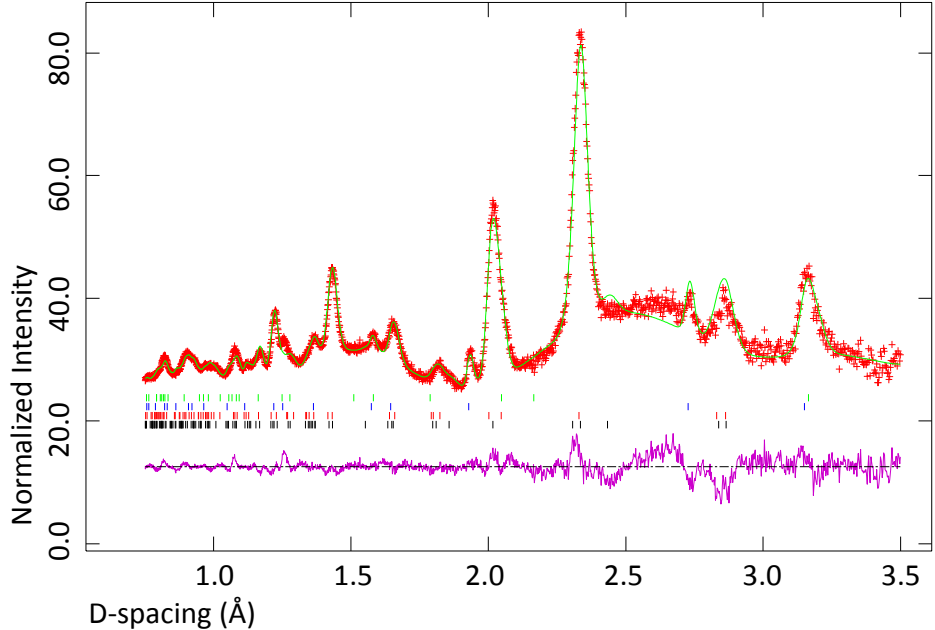}
\caption{\label{Figure2}Observed (red) and calculated (green) time-of-flight neutron powder diffraction data and its difference curve between measured and computed curves (purple) for a Pb(Zr$_{0.54}$Ti$_{0.46}$)O$_3$ sample at 303 K and 3 GPa. The tick marks, from down to up, are from the $R3c$, $Cm$, NaCl (pressure standard) and graphite (from the pressure chamber) phases. The statistical figures of merit were: $\chi^2=2.300$, $R_{wp}=2.04$ \%, $R_p = 1.42$ \% and the background substracted $R$ parameters were $R_{bwp}=2.75$ \% and $R_{bp}=1.65$ \%.}
\end{center}
\end{figure}

\begin{figure}[htb!]
\begin{center}
\includegraphics[angle=0,width=10cm]{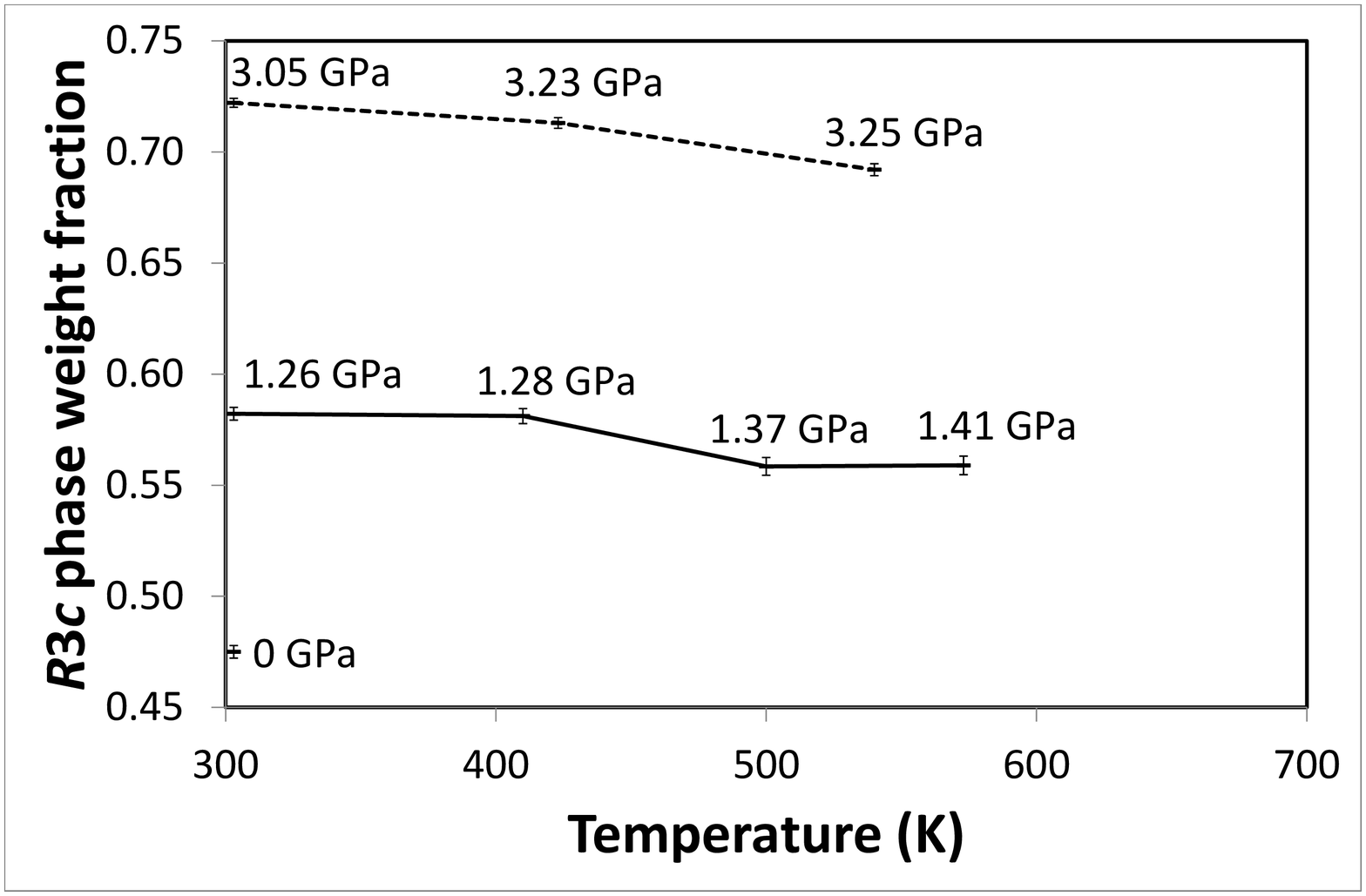}
\caption{\label{Figure3}Rhombohedral weight fraction at ambient conditions and as a function of temperature at approximately 1 and 3 GPa pressures.}
\end{center}
\end{figure}

\begin{figure}[htb!]
\begin{center}
\includegraphics[angle=0,width=13cm]{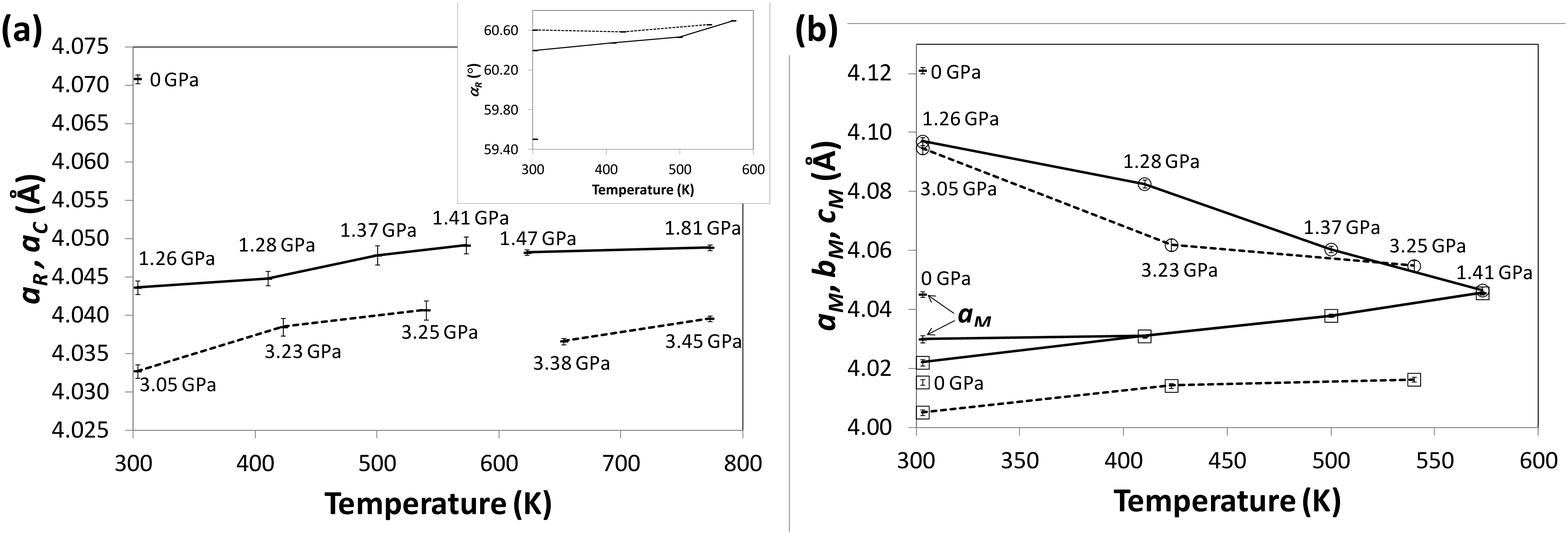}
\caption{\label{Figure4}Lattice parameters of the $R3c$, $Cm$, $P4mm$ and $Pm\bar{3}m$ phases at ambient conditions and as a function of temperature at approximately 1 and 3 GPa pressures. Monoclinic and tetragonal $b_M$ axis lengths are surrounded by a square. The $c_M$-axis values are enclosed by a circle. The $Cm$ phase transformed to the $P4mm$ phase at around 400 K at 1 GPa pressure. At ambient conditions the monoclinic angle $\beta$ was $90.01(97)^{\circ}$ and at 1.26 GPa pressure $\beta$ was $90.62(4)^{\circ}$. The 3 GPa data is indicated by a dotted line. Due to the thermal pressure, the pressure values of the highest two temperatures (cubic phase) are larger. The inset shows the rhombohedral angle $\alpha$.}
\end{center}
\end{figure}

\begin{figure}[h!]
\begin{center}
\includegraphics[angle=0,width=9cm]{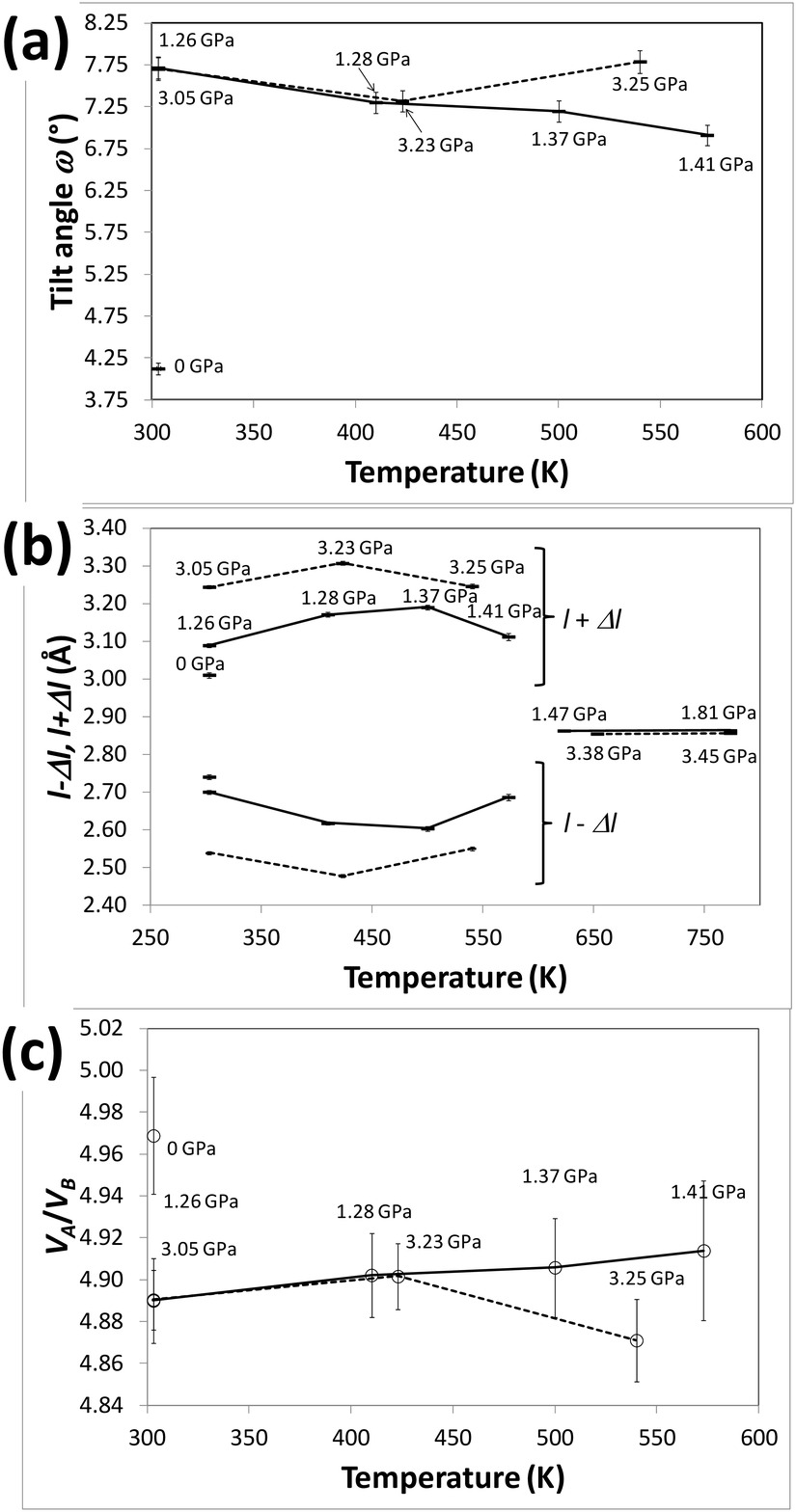}
\caption{\label{Figure5} Octahedral tilt angles (a), octahedral edge lengths (b) and  polyhedral volume fractions of the $R3c$ phase at ambient conditions and as a function of temperature at approximately 1 and 3 GPa pressures. }
\end{center}
\end{figure}

\begin{figure}
\begin{center}
\includegraphics[angle=0,width=10cm]{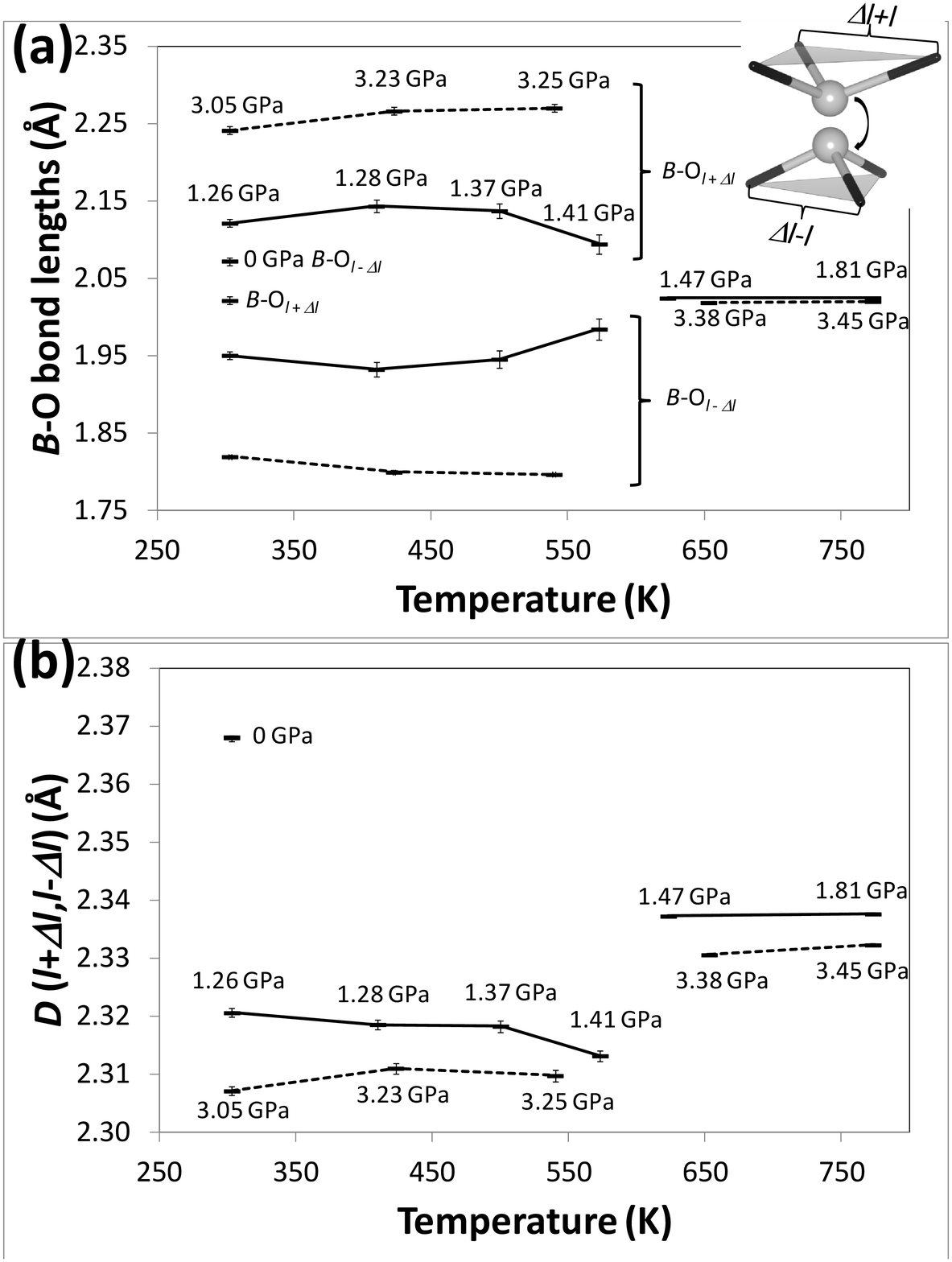}
\caption{\label{Figure6}(a) $B$-cation (Zr or Ti) and oxygen bond lengths in the rhombohedral phase. The difference between $B-\mathrm{O}_{\Delta l+l}$ and $B-\mathrm{O}_{\Delta l-l}$ bond lengths increases with increasing pressure. The decrease in difference seen at 1.41 GPa pressure is probably related to the vicity of the transition to the cubic phase. (b) The distance $D(l+\Delta l, l-\Delta l)$ between the oxygen triangles. In both panels,  the 3 GPa data are indicated by dotted lines. The inset shows the displacement of the $B$ cations under pressure. At ambient pressures the $B$ is closer to the larger triangle and displaces towards smaller triangle under pressure.}
\end{center}
\end{figure}

\paragraph{Structural model.}The X-ray diffraction pattern collected on Pb(Zr$_{0.54}$Ti$_{0.46}$)O$_3$ powder is characteristic to the morphotropic phase boundary composition, the most apparent indication of a two-phase co-existence is seen from the pseudo-cubic 200-reflections. Thus, the $R3c+Cm$ structural model (see refs. \onlinecite{Frantti_JPCM_2003,Yokota,Phelan}) was used for the refinements of the low-pressure data at ambient temperatures. Refinements indicated that the monoclinic distortion continuosly vanished with increasing hydrostatic pressure and increasing temperature. The monoclinic structure became tetragonal and was correspondingly modelled by the $P4mm$ space group. Fig. \ref{Figure2} shows the pattern collected at 3 GPa pressure at room temperature and the computed intensity. 
At ambient conditions the majority phase was monoclinic, see the 0 GPa datum in Fig. \ref{Figure3}. With increasing pressure the situation changed significantly (Fig. \ref{Figure3}), accompanied by drastic changes in rhombohedral tilts and polyhedral volumes (Fig. \ref{Figure4}). Slight increase of the tetragonal phase fraction with increasing temperature at constant pressure is seen in Fig. \ref{Figure3}. The lattice parameters given in Fig. \ref{Figure4} indicate that the $Cm$ phase does not continuously transform to the rhombohedral phase: the difference between the rhombohedral and monoclinic structures remains large up to the point at which the $Cm$ phase  continuously transforms to $P4mm$ phase. Instead, through the studied pressure and temperature range yet there are significant changes in the phase fractions. This is in line with the first-order phase transition and shows that no continuous polarization rotation occurs.
Thus, the phase stabilities as a function of pressure and temperature follow well the predictions based on the first-principles studies carried out for PbTiO$_3$ \cite{Frantti_JPCB_2007,Frantti_JPCB_2009}. Further, the entropy term seems to have a crucial role for setting the boundary between the pseudo-tetragonal and rhombohedral phases: the pseudo-tetragonal phase fraction increases with increasing temperature at constant pressure. 
\paragraph{Octahedral tilting.}Figure \ref{Figure5} shows the octahedral tilts in the $R3c$ phase and the two characteristic octahedral edge lengths, $l-\Delta l$ and $l+\Delta l$. The octahedral tilt increases with increasing pressure, though the tilt angle saturates at high pressures.
Thus with increasing pressure the volume fraction of the octahedra increases, consistently with the idea that, when compared to the tightly filled oxygen octahedra, lead ions have excessive space inside cuboctahedra formed from 12 oxygen atoms. In addition to the oxygen octahedral tilting also another 
mechanism can be seen:  the continuous expansion of the $l+\Delta l$ and contraction of the $l-\Delta l$. Fig. \ref{Figure6} (a) shows the $B$-cation (Zr or Ti) and oxygen bond lengths in the rhombohedral phase. At ambient conditions the $B$ cations are closer to the larger oxygen triangle, consistently with the earlier data \cite{Frantti_JPCM_2003}. This situation changes with increasing pressure: it is seen that the $B$-cations are 
closer to the small oxygen triangle, indicating that at higher pressures the $B$-cations favour to form a small tetrahedron rather than being centered closer to the octahedron center, see the inset of Fig. \ref{Figure6}. Positions in which the $B$ cations are closer to the large triangle is clearly unfavourable as it would result in bond lengths failing to fullfill the bond-valence criteria.  At 3 GPa pressure the distance between the vertex of the large oxygen triangle and triangle center alone is slightly larger than the given $B$-O lengths. 
For piezoelectricity this has important consequences: if stress is sufficiently strong, it switches the position of the $B$ cations from a larger oxygen triangle towards the smaller oxygen triangle thus contributing to the  intrinsic piezoelectricity. Thin film technology allows a deposition of selected crystal planes in which the biaxial stress can be adjusted by choosing the substrate and composition so that the piezoelectric proeprties can be optimized.

Fig. \ref{Figure6} (b) gives the distance between the oxygen triangles, $D(l+\Delta l, l-\Delta l)$. 
Figs. \ref{Figure5}(b) and 
\ref{Figure6} (b) show that whereas $D(l+\Delta l, l-\Delta l$) and $ l-\Delta l$ both decrease and $ l+\Delta l$ increases 
significantly when pressure increases from 0 to 1 GPa, $D(l+\Delta l, l-\Delta l)$ hardly changes when pressure increases from 
1 GPa to 3 GPa. Instead, $ l-\Delta l$ and $ l+\Delta l$ decrease and increase significantly, respectively.
\paragraph{Reversibility.}A first-order transition is frequently characterized by a two-phase co-existence region of metastable and stable phases as a function of the thermodynamic variable (e.g., temperature or pressure). In piezoelectric materials this is one source of irreversibility (other significant contribution being due to the irreversible domain wall motion). It is interesting to note that the recovery run, carried out after the high-pressure and high-temperature cycles, revealed that the rhombohedral phase fraction had increased when compared to the prior the high-pressure situation. This suggests that high-pressure synthesis is a useful way to prepare single-phase rhombohedral ceramics in the vicinity of the MPB. The advantage over the Zr-rich rhombohedral ceramics is that in the vicinity of the phase transition certain piezoelectric constants are more susceptible to external stimuli. We note that recent neutron powder \cite{Yokota} and single crystal \cite{Phelan} diffraction studies revealed that there is a secondary monoclinic $Cm$ phase present in the Zr-rich case, together with the rhombohedral $R3m/R3c$ phases. Recent single crystal study also showed that the diffraction data, collected on Pb(Zr$_{0.54}$Ti$_{0.46}$)O$_3$ and  Pb(Zr$_{0.69}$Ti$_{0.31}$)O$_3$ samples are better interpreted in terms of the rhombohedral and  monoclinic phases, rather than by the adaptive phase model \cite{Gorfman}. The two-phase co-existence and the nature of the phase transition are believed to be crucial for the piezoelectric properties.

\section{Conclusions}
High-pressure neutron powder diffraction experiments were applied to the classical piezoelectric compound, Pb(Zr$_{0.54}$Ti$_{0.46}$)O$_3$. Weight fraction changes between the rhombohedral $R3c$ and monoclinic $Cm$ (low-pressures and room temperature) or between tetragonal $P4mm$ phases as a function of hydrostatic pressure and function were determined. The $Cm$ phase was observed only at low-pressures and ambient temperatures as it transformed to the $P4mm$ phase at approximately 1 GPa and 400 K. As the earlier computations predicted, the rhombohedral phase was favored at higher pressures, whereas the added heat increased the monoclinic phase fraction at constant pressure. This largely contributes to the extrinsic piezoelectricity. These findings are in line with the computational model according to which the phase boundary between the rhombohedral and tetragonal phase in pressure-temperature plane is dictated by the two competing terms, octahedral tilting and entropy term. No support for a continuous polarization rotation was found. The oxygen octahedra was significantly distorted under pressure, accompanied by a significant displacement of the $B$ cations. This contributes to the intrinsic piezoelectricity. After the experiments the fraction of the $R3c$ phase was larger than initially, suggesting that the $Cm$ phase is not stable. This is consistent with the first-order phase transition $Cm\rightarrow R3c$.

\section*{Acknowledgements}
The research work was supported by the collaboration project between the Center of Excellence for Advanced Materials Research at King Abdulaziz University in Saudi Arabia and the Aalto University and the Academy of Finland (Projects 207071, 207501, 214131, and the Center of Excellence Program 2006-2011). This work has benefited from the use of the Lujan Neutron Scattering Center at Los Alamos Neutron Science Center, which is funded by the U.S. Department of Energy's Office of Basic Energy Sciences. Los Alamos National Laboratory is operated by Los Alamos National Security LLC under DOE contract DE-AC52-06NA25396.

\end{document}